\documentclass[conference]{IEEEtran}
\IEEEoverridecommandlockouts
\usepackage{cite}
\usepackage{amsmath,amssymb,amsfonts}
\usepackage{algorithmic}
\usepackage{graphicx}
\usepackage{textcomp}
\usepackage{xcolor}
\usepackage[]{float}
\usepackage{hyperref}
\def\BibTeX{{\rm B\kern-.05em{\sc i\kern-.025em b}\kern-.08em
    T\kern-.1667em\lower.7ex\hbox{E}\kern-.125emX}}
\begin{document}

\title{Using historical utility outage data to compute overall transmission grid resilience
\thanks{The authors gratefully acknowledge funding in part from NSF grants ECCS-1254549, CNS-1735513, DGE-1144388, ECCS-1609080, and CNS-1735354.}
}

\author{\IEEEauthorblockN{Molly Rose Kelly-Gorham}
\IEEEauthorblockA{Electrical Engineering \\
University of Vermont\\
Burlington VT, US \\
mkellygo@uvm.edu}
\and
\IEEEauthorblockN{Paul Hines}
\IEEEauthorblockA{Electrical Engineering \\
University of Vermont\\
Burlington VT, US \\
Paul.Hines@uvm.edu}
\and
\IEEEauthorblockN{Ian Dobson}
\IEEEauthorblockA{Electrical and Computer Engineering \\
Iowa State University\\
Ames IA, US \\
dobson@iastate.edu}
}

\maketitle

\begin{abstract}
Given increasing risk from climate-induced natural hazards, there is growing interest in the development of methods that can quantitatively measure resilience in power systems.
This work quantifies resilience in electric power transmission networks in a new and comprehensive way that can represent the multiple processes of resilience.
A novel aspect of this approach is the use of empirical data to develop the probability distributions that drive the model.
This paper demonstrates the approach by measuring the impact of one potential improvement to a power system.
Specifically, we measure the impact of additional distributed generation on power system resilience.
\end{abstract}

\begin{IEEEkeywords}
resilience, empirical data, blackout, risk
\end{IEEEkeywords}

\section{Introduction}

While there is ongoing debate about exactly how to define resilience in electric power systems, within the field of ecology, resilience is well defined.
According to~\cite{walker2004resilience}, ``resilience is the capacity of a system to absorb disturbance and reorganize while undergoing change so as to still retain essentially the same function, structure, identity, and feedbacks.''
In order to understand how resilient a system is, therefore, we need to understand both
the susceptibility of that system to failures,
and the ability of that system to recover from failures.
In this paper we propose a method for quantitatively measuring resilience in electric power systems based on modeling (using empirical data)
five key processes:
\begin{enumerate}
\item Stress leads to component outages or failures
\item Cascading failures propagate within the network, causing additional outages
\item Services are interrupted (load is shed)
\item Restoration of failed components and unserved load
\item Quantification of blackout impacts
\end{enumerate}
All large resilience events involve these five processes to some extent. For example, in the August 14, 2003 blackout:
(1) lines sagged into trees,
(2) outaged lines triggered a cascade of failures,
(3) around 50 million people were left without power,
(4) power was incrementally restored over a period of hours to days,
(5) estimates of direct blackout cost ranged from 4 to 12 billion dollars.

\begin{figure}
    \centering
    \includegraphics[width=0.85\columnwidth]{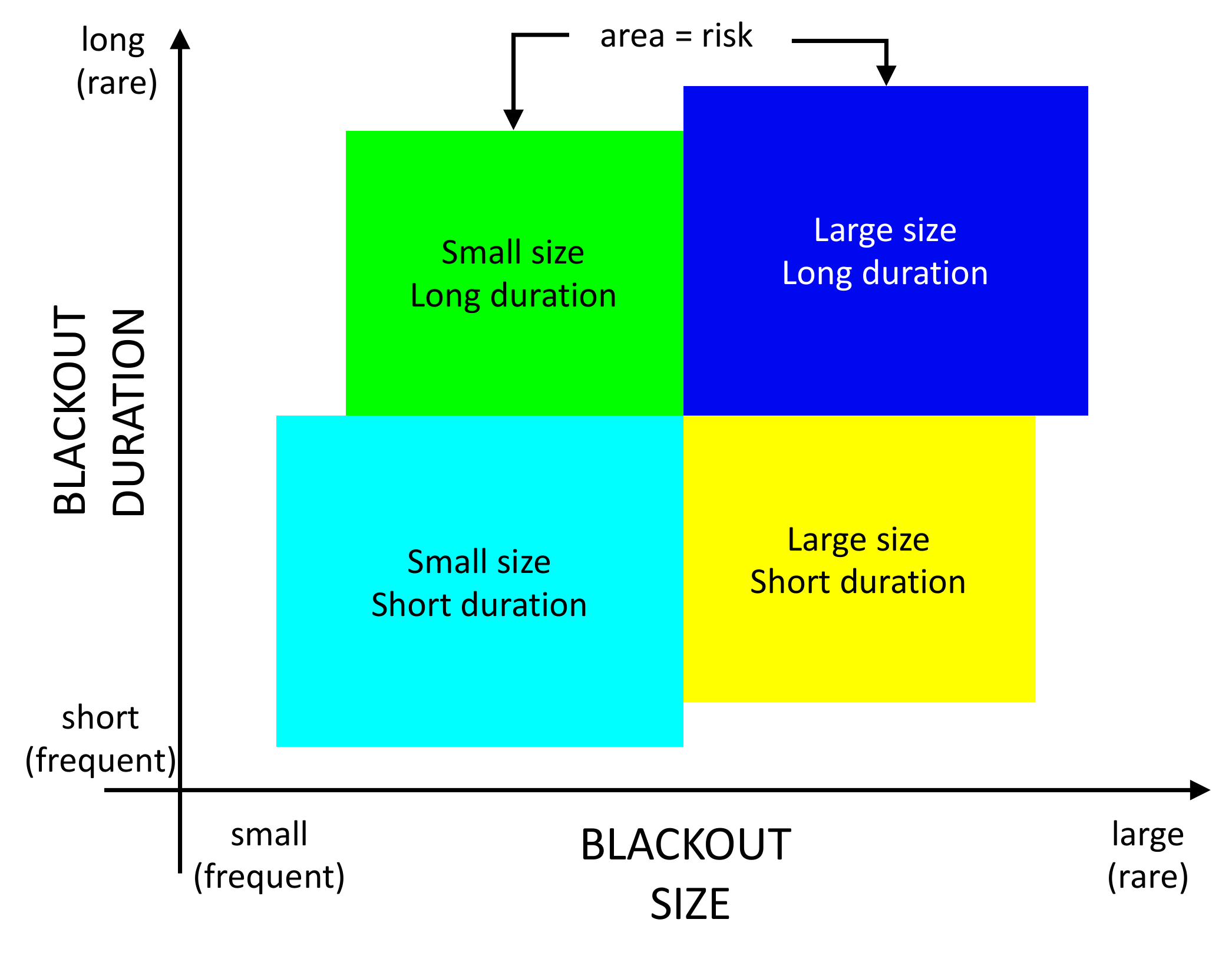}
    \vspace{-0.15in}
    \caption{Visualizing the relative contribution to resilience risk from blackouts of different sizes and durations. The area of each rectangle is proportional to the blackout risk in each category. This figure separates blackouts along two dimensions: blackout size (load lost, customers unserved) and blackout duration. Size and duration inversely correlate with probability: Small (or short) blackout are relatively frequent and large (or long) blackouts are rare. Because risk is the product of probability and impact, the risk from blackouts of different sizes is frequently similar.}
    \label{fig:event-sizes}
\end{figure}

The most effective analyses of a particular resilience event, such as a blackout, measure the event's impact or social cost in a number of dimensions.
Two of the most important dimensions for understanding resilience are the size of the blackout (usually the initial extent in customers or power shed) and the amount of time required to restore the load (the duration).
Some events, like the European blackout of Nov.~4, 2006 have enormous impact in geographic size (15 million customers), but are relatively short in duration (hours).
Other events have smaller geographic size (e.g., the impact of Hurricane Irma on Puerto Rico), but the restoration process takes a very long time (almost 1 year until complete restoration).
In order to understand resilience, we need to be able to quantify the risk that comes from events in each of these dimensions, since it is likely that different strategies are needed to deal with these different types of risks.
Figure~\ref{fig:event-sizes} provides one way to think about these two different dimensions.

Due to the critical importance of designing civil infrastructure to withstand extreme events, there is substantial research on infrastructure resilience in the civil engineering literature.
Several have studied the impact of weather events on power systems infrastructure~\cite{liu2005negative,han2009estimating,Bruneau:Resilience-earthquake}.
Others have developed sophisticated models of interdependence among infrastructure systems~\cite{satumtira2010synthesis,reed17,wu2016modeling}.
However, there is often a need for additional power systems detail in order to fully understand resilience in electric energy systems.

In power systems there is long history of valuable research into many elements of the resilience problem.
Many have studied the risk of exogenous events, such as storms~\cite{Dunn18} on the failure of power systems components.
Others study the impact of multiple contingencies on power systems (e.g.,~\cite{MultiElementContScreen:Davis, MassiveContAnal:Huang,PerfIndex:Ejebe,MLS:Carleton}).
There is a growing body of literature regarding the processes and risk associated with how initiating outages trigger cascading failures and thus large blackouts~\cite{IanCascade,Clarfeld2019,song_dynamic_2016}.
Power system restoration has been an important topic for years~\cite{Adibi94}.
Motivated at least in part by recent severe resilience events, such as Superstorm Sandy~\cite{ji2016large}, Hurricane Katrina~\cite{kwasinski2009telecommunications}, and the Chilean earthquake of 2010~\cite{araneda2010lessons}, many have proposed new methods for improving power systems restoration processes for transmission networks~\cite{coffrin15} and for distribution networks~\cite{Tan2017}.

Initial efforts within the power systems community have developed metrics for measuring resilience of the power grid~\cite{Panteli:MetricsResilience}. A few have combined many of the five processes and initial work on the remaining processes into a framework to measure the resilience~\cite{Panteli:Fragil-Prob-Adapt,Tan2018,Nan2017}.
Others~\cite{Panteli:Fragil-Prob-Adapt} used structural analysis techniques to provide the fragility curve for transmission towers and measure the effect of time and space variation in stress from natural hazards.
Ref.~\cite{Tan2018} 
used optimal restoration in combination with system hardening to optimize the components to harden.
Nan et al.~\cite{Nan2017} systematically incorporated the components of resilience into one metric, which was combined with performance measures, and this paper also included in the implementation steps an approach to model interactions between critical infrastructures. Methods to model the effects of multiple hazards together have been explored and results compared with distribution network outage data in~\cite{Alvehag:Reliability-Distrib-Weather}.

But in our estimation there are two key gaps in the existing literature.
First, there is a need for additional work that models the resilience problem as a whole, providing methods for quantitatively (and tractably) modeling each of the five processes: Stress, Cascading, Failure, Recovery, and Analysis.
Including all five resilience processes is critical for addressing important policy questions, such as `what types of resources or investments provide the most resilience value?'

Second, there is a need for resilience-modeling methods that are driven by real data.
Obtaining sufficiently detailed utility data is difficult, but some real data is becoming available and is even publicly available~\cite{DobsonHICSS18}. Real-world statistical data describing the processes of resilience offer a strong opportunity to form a comprehensive resilience model that is directly driven by this data, which can mitigate many of the challenges associated with the tricky modeling assumptions that are required when real data are not available.
This paper describes and demonstrates a platform or framework for doing just this: using real-world data from bulk power systems to comprehensively and quantitatively model all the processes of resilience.

In this work, we aim to evaluate resilience comprehensively and simply by approximating the multiple processes of resilience with probability distributions driven by real data.
We call this data-driven approach CRISP, standing for Computing Resilience Interactions Simulation Platform.
To illustrate the use of CRISP,
 we demonstrate how  it can be used to address the significant policy question of  quantitatively assessing the extent to which distributed generation can enhance resilience in a power system test case.

 \section{Overview of CRISP model}

This section provides a high-level overview of how the CRISP model proceeds through the five resilience processes.
One thing to note is that this implementation of CRISP assumes that the resilience processes do not overlap.
That is, we make the reasonable idealization that the five processes of resilience occur separately and in sequence.

\subsection{Outages after cascading}

In this implementation of CRISP we use real data to model the first two processes of resilience (Stress and Cascading).
Specifically, we use historical data to estimate the distribution of the number of outages that occur in close temporal proximity, given that one or more transmission outages occur. Here we use the term ``cascade" to refer to a group of line outages in close temporal proximity \cite{DobsonPS12}.

The annual rate of cascades of outages occurring is $c$.
Given that a cascade of outages has occurred,
the total number of components outaged after cascading $N_L$ includes the initial outages as well as the outages that result from any cascade propagation that occurs after the initial outages.
Due to the different time scales of cascading and restoration, this assumption should make a minor impact on the measured impact of the blackout. 
The total number of line outages  $N_L$ is modeled as a Zipf distribution
with parameter $s$:
\begin{align}
    N_L&\sim {\rm Zipf}(s)\notag\\
    P[N_L=n_L]&=\frac{ 1}{\zeta(s)} \frac{ 1}{n_L^s} \quad,n_L=1,2,3,...
     \label{zipf}
\end{align}
The parameter $s>0$ controls the slope of the Zipf distribution on a log-log plot.
A smaller $s$ gives a more frequent large number of outages and corresponds to a transmission system with more stress, such as worse weather or peak loading.

Given the number of lines outaged, the individual lines that go out are chosen randomly.
Future versions of the model will adjust the relative outage rates from historical outage data and network structure.

Note that we do not describe the evolution in time or the details of the component stress or outages or cascading, we simply rely on observed data for the statistics of the outcome of these processes.
This substantially reduces the number of modeling assumptions we need to make in our quantification of resilience. 

\subsection{Blackout}
Given the individual line and generator outages, we need to evaluate which loads are shed as a result. This is done by running an optimal power flow called LS-OPF that allows load shedding for the grid that remains intact. 
In particular, the load shed is minimized subject to the DC load flow equations, line flow limits, generation limits, and load limits. The total load shed is denoted as $L_0$.


\subsection{Restoration}
Given the lines and generators that are out of service at the end of the outage process, CRISP next models the time $R_k$ after the end of the outage process at which component $k$ is restored.
The component restore times $R_k$ are independent samples from the distribution
\begin{align}
    R_k&\sim {\rm LogNormal}(\mu,\sigma^2)\notag\\
    f_R(r)&=\frac{1}{r\sqrt{2\pi}\,\sigma}\exp\Big[\frac{-(\ln{r}-\mu)^2}{2\sigma^2}\Big]\quad,r>0
    \label{lognormal}
\end{align}

The restoration process is evaluated at discrete times $t_1,t_2$, ..., $t_n$  after the end of the outages.
The restoration process starts at time $t_0$ with the
lines and generators that are outaged at the end of the outage process.
At each time $t_k$, the components restored in the time interval $[t_{k-1},t_k]$ are determined from their individual restoration times. Then the LS-OPF is run to evaluate which loads are shed at time $t_k$.
The total load not served at time $t_k$ is denoted by $L_k$.
The restoration process continues until all outaged components are restored.

\subsection{Blackout impact}
Next, we quantify the blackout size by the energy unserved $S$, which is obtained from the load shed during the restoration process as
\begin{align}
    S=\sum_{k=1}^n(t_k-t_{k-1})L_{k-1}
\end{align}
Since the CRISP model is generated stochastically by sampling from probability distributions,
each pass through the model gives a sample outcome, and many such samples are accumulated to give an empirical distribution of energy unserved $S$.
Each sample of $S$ is the energy unserved in one cascade of outages.
If the annual expected energy unserved over a time period is desired, then this
is the mean of the distribution of $S$ multiplied by the annual cascade rate $c$.
\begin{align}
    \mathrm{EENS} = c  \mathrm{E}[S]
\end{align}

\section{Utility data driving the CRISP model}

To illustrate the CRISP modeling, we use statistical data from  published detailed outage data of a large transmission grid in the Northwest USA from 1999 to 2012~\cite{BPAwebsite}.
This detailed historical outage data has been processed for other uses in previous work~\cite{
DobsonPS12, DobsonPS16,KancherlaPS18}.

North American utilities and many utilities world-wide routinely collect similar data, and it is intended that CRISP modeling would be applied in practice with each utility's own detailed outage data that they have already collected together with the DC load flow model of that utility's transmission grid.
It is convenient (but not essential to the approach) to fit the empirical probability distributions from the data with standard distributions that are then sampled to drive the CRISP model, instead of directly sampling from the empirical distributions themselves.

\subsection{Number of lines outaged after a cascade}

We need the distribution of the total number of transmission lines outaged at the end of a cascade
(that is, both the initial and subsequent cascading outages).
We use the automatic line outages in the data that are grouped into cascades if they occur sufficiently close in time using the simple processing of~\cite{DobsonPS12}.
To ensure that the statistics reflect successful protection actions as well as the rarer events involving a succession of multiple outages, the cascades are defined to also include single or only a few outages that do not propagate further.

\begin{figure}[H]
\centering
\includegraphics[width=\columnwidth]{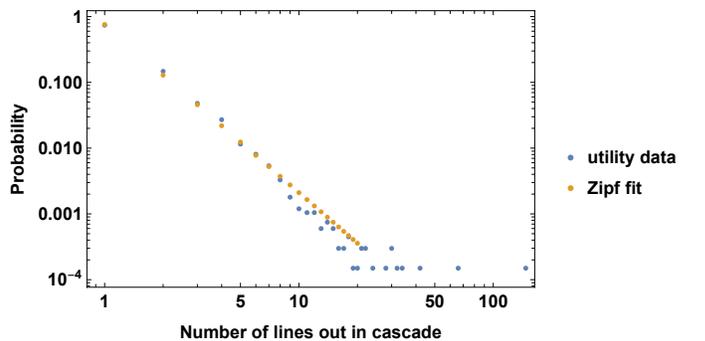}
\vspace{-.2in}
\caption{Probability distribution of the number of lines outaged in a cascade on  log-log plot showing empirical distribution from utility data and fit with Zipf distribution.}
\label{distnumberoflinesCRISP}
\end{figure}

The empirical distribution of the number of line outages in a cascade is shown in Fig.~\ref{distnumberoflinesCRISP}.
This empirical distribution is fit with the Zipf distribution (\ref{zipf}) shown in Fig.~\ref{distnumberoflinesCRISP} using the formula in~\cite{ClausetSIAM09}, which gives the Zipf parameter $s=2.56$.

\subsection{Transmission line restoration time}

The empirical distribution of the transmission line restoration time is obtained directly from the utility data~\cite{BPAwebsite} and fit with the lognormal distribution (\ref{lognormal}) as in ~\cite{KancherlaPS18}.
The fitted lognormal parameters are $\mu=3.66$ and $\sigma=2.43$.

\subsection{Power system case data}
We edited the 39 bus case in MATPOWER to be $n-1$ secure, by adding redundant lines and increasing line limits where necessary. As an experiment we added distributed generation to the load buses in the original and $n-1$ secure 39 bus case to show the positive effect of well managed distributed resources on network resilience.

\section{CRISP Implementation Details}
This section describes the modeling and implementation details we are using for this work, but which could be done differently depending on what questions motivate the analysis.

\subsection{Data driven approach to line outages}

The number of lines that outage and their restoration times are sampled
from their respective distributions (\ref{zipf}) and (\ref{lognormal}), using standard inverse transform sampling that samples from a uniform distribution between 0 and 1 and then applies the inverse function of the CDF. For this initial work, the particular lines outaged are chosen randomly.

\subsection{Find initial load shed}
Step 2 finds the initial load shed that is necessary to stabilize the system directly after the outages found in step 1 occur. The formulation of the load shedding OPF (LS-OPF) problem is as follows:
\begin{align}
    \max &\quad \sum \Delta P_d \\
    \mathrm{s.t.} &\quad B \Delta \theta  = \Delta P_g - \Delta P_d \\
    &\quad -\overline{P_{ft}} \leq P_{ft}[0] + \frac{1}{X_{ft}} \Delta \theta_{ft} \leq \overline{P_{ft}} \\
    &\quad \underline{\Delta P_g} \leq \Delta P_g \leq \overline{\Delta P_g} \\
    &\quad \underline{\Delta P_d} \leq \Delta P_d \leq \overline{\Delta P_d}
\end{align}
where $\Delta P_d$ is the total load shed and is negative, $B$ is the susceptance matrix, $\Delta \theta$ is the change in voltage angles at each bus, $\Delta P_g$ is the change in generation at each bus, $\overline{P_{ft}}$ is the maximum power flow over the line from bus $f$ to bus $t$, $P_{ft}[0]$ is the initial power flow over the line, $X_{ft}$ is the reactance of the line between bus $f$ and bus $t$, $\Delta \theta_{ft}$ is the change in the difference between the voltage angles at bus $f$ and bus $t$, $\underline{\Delta P_g}$ is the maximum possible decrease in generation, $\overline{\Delta P_g}$ is the maximum possible increase in generation, $\underline{\Delta P_d}$ is the maximum decrease in demand, which should be the negative of the amount of demand, and $\overline{\Delta P_d}$ is the maximum increase in demand, which initially will be 0.

    If the outages create islands in the system, we first find each island and we solve for the unserved load separately and find the sum.

\subsection{Data driven approach to restoration process}
The model of the restoration process loops through the repair times of each line in temporal order, and at each repair time, $t_k$, finds the optimal load shed, $L(t_k)$.
We use a similar optimal power flow to the LS-OPF with updated bounds on the variables $\Delta P_d$ and $\Delta P_g$ depending on the current state of $P_d$ and $P_g$.
We do not allow line switching in the optimization.

\subsection{Compute energy unserved}

\begin{figure}
    \centering
    \includegraphics[trim=0cm 0cm 0cm 1cm, clip=true,width=0.9\columnwidth]{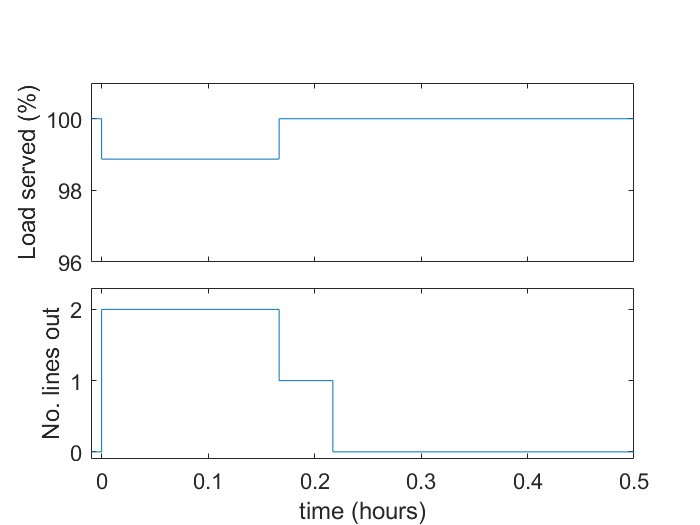}
    \vspace{-.1in}
    \caption{Two line event on the 6 bus test case where some initial load shed is necessary, and full service is restored after one line is restored.}
    \label{fig:case6ww-single}
    \vspace{.1in}
    \includegraphics[trim=0cm 0cm 0cm 1cm, clip=true, width=0.9\columnwidth]{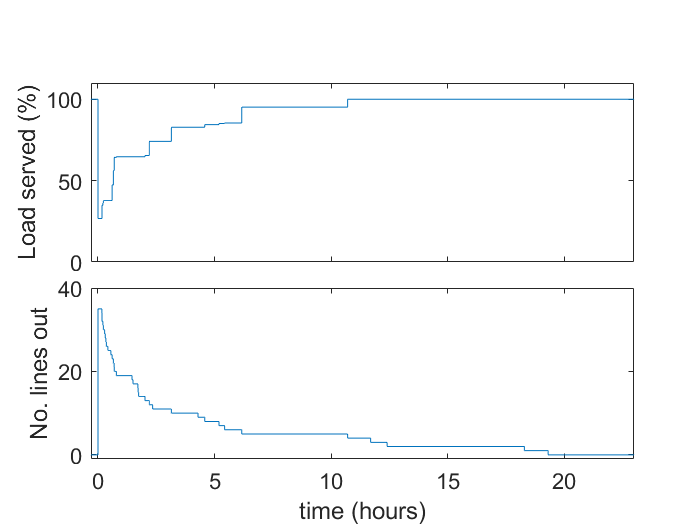}
    \vspace{-.1in}
    \caption{35 line event on the 39 bus test case on top is the percent of load served and on the bottom is the number of lines out over time.}
    \label{fig:case39-severe}
\end{figure}

We integrate the unserved load to measure the area formed by the horizontal line of 100\% load and the load served curves in the top plots of figures \ref{fig:case6ww-single} and \ref{fig:case39-severe}. This area is the energy not served, ENS, for this event.\\

This implementation of the CRISP model will be made publicly available at \url{https://github.com/phines/infrastructure-risk/MEPS}. Figure \ref{fig:case6ww-single} shows the result of a two line event on the 6 bus MATPOWER test case at 0 hours in terms of the percent of load served and the number of lines out over time for the full restoration process. Fig.~\ref{fig:case39-severe} shows the percent of load served and the number of lines out over time for an extreme 35 line event on the original 39 bus MATPOWER test case.

Since our model is stochastic, and the effect of one event on a network is not a good measure of its resilience, we simulate 10000 events to find the distribution of outcomes.
CRISP measures the energy not served (ENS) of each blackout in units of MWh.
Note that many of the initiating outages lead to 0 MWh of unserved energy.

\section{39-bus system results from CRISP}

In order to demonstrate CRISP, we applied the model to the IEEE 39 bus test case.
In order to find the distribution of ENS, we measured ENS over 10,000 model runs, and then compared six different variants of the test case designed to address two important questions.
First, how does resilience change as more distributed generation (DG) is added to the test case?
Second, what is resilience impact of adjusting the test case to make it $n-1$ secure?

\begin{figure}
    \centering
    \includegraphics[trim=0cm 0cm 0cm 0cm, clip=true, width=\columnwidth]{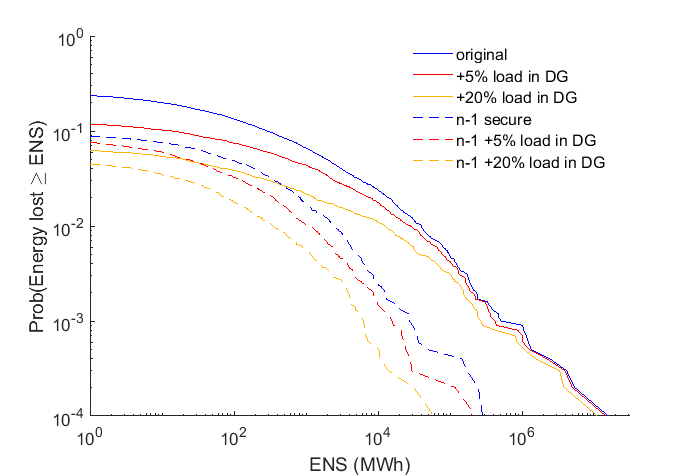}
    \vspace{-0.2in}
    \caption{Complementary cumulative distribution of energy not served, assuming an outage, from 10,000 samples of CRISP simulation on the 39 bus test case.  Percent of load that is distributed generation is varied for the original 39 bus system and for the system upgraded to be $n-1$ secure. The probability of no load shed for the original cases is approximately 75\% (blue), 87\% (red), and 94\% (yellow), and for the and $n-1$ cases is approximately 91 \% (blue dashed), 94 \% (red dashed), and 97 \% (yellow dashed).
    }
    \label{fig:case39-exp1}
\end{figure}

In order to address the first question (DG), we added new distributed generators at each load bus with capacity equal to 5\%, and 20\% of the base case load, and then computed the ENS distribution.
In order to address the second question, we modified the 39 bus case to be $n-1$ secure by adding 8 redundant parallel lines, and increasing line power flow ratings where necessary.
The results from 10,000 runs on each of the resulting six cases are shown in Fig.~\ref{fig:case39-exp1}.

Figure \ref{fig:case39-exp1} clearly shows that the added dispatchable DG substantially increases the resilience of the network.
The DG leads to a particularly large reduction in risk from small events for the original 39 bus case, bringing the number of events with 0 ENS from 75\% to 87\% with only 5\% DG and even further to 94\% with 20\% DG.
The DG had a similar effect on the $n-1$ secure version of the 39 bus case, however the added 5\% DG has a less pronounced impact on risk, relative to the impact of 5\% DG on the original case.
$n-1$ security already substantially reducing the number of small blackouts.

\begin{figure}
    \centering
    \includegraphics[trim=0cm 0cm 0cm 0.5cm, clip=true,width=\columnwidth]{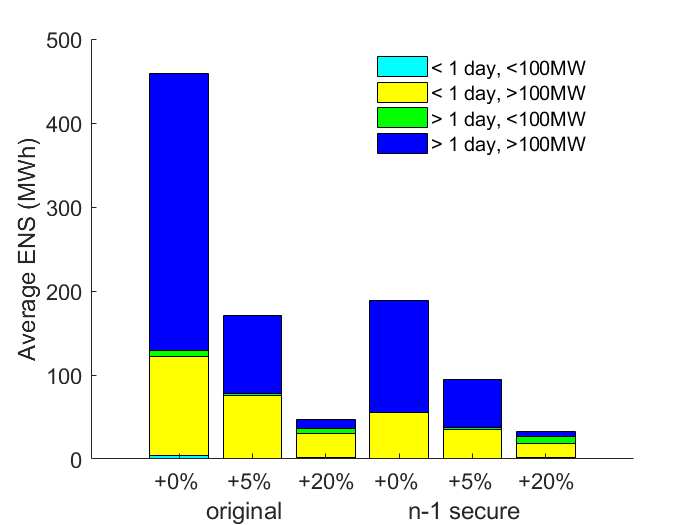}
    \vspace{-0.1in}
    \caption{Each bar shows average energy unserved per event separated into 4 categories based on the recovery time to 99.9\% of the load served and the initial load shed for 39 bus test case. Percent of load that is distributed generation is varied for the original 39 bus system and for the system upgraded to be $n-1$ secure.}
    \label{fig:case39-bar}
\end{figure}

\begin{figure}
    \centering
    \includegraphics[trim=1cm 0.5cm 1cm 1cm, clip=true, width=\columnwidth]{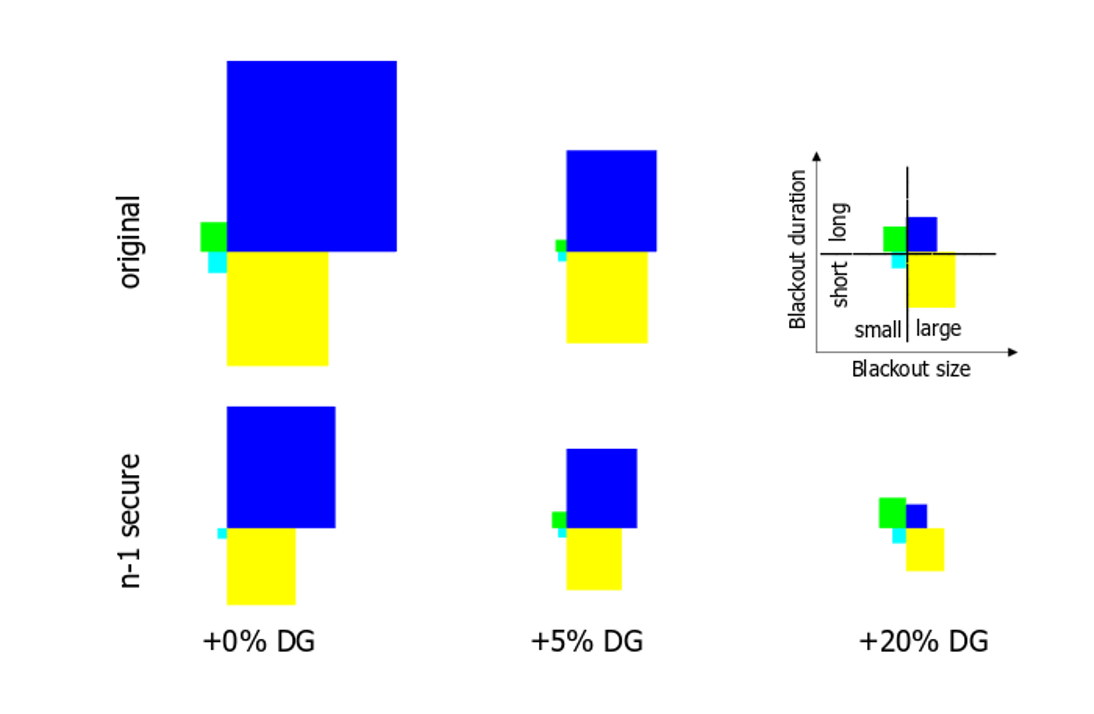}
    \vspace{-0.3in}
    \caption{The relative risk, shown as relative area, of four categories of events based on the recovery time to 99.9\% of the load served and the initial load shed. 
    Percent of load that is distributed generation is varied for the original 39 bus system and for the system upgraded to be $n-1$ secure. The colors also correspond with figure \ref{fig:case39-bar}.}
    \label{fig:case39-squares}
\end{figure}
In Figs.~\ref{fig:case39-bar} and \ref{fig:case39-squares} we separate the events into 4 categories to separately analyze the risk of events that have different sizes and durations.
A measure of the relative risk of the events within the categories is:
\begin{align}
r &= \frac{1}{n} \sum_{i \in E_k} \mathrm{ENS}_i
\end{align}
where $r$ is the relative risk of the category, $n$ is the total number of events, and $E_k$ is set of events within category $k$.
This is shown in the stacked bar chart in figure \ref{fig:case39-bar} and represented in the area shown in figure \ref{fig:case39-squares}.
The 4 categories separate the small and large events and the short and long events (see Fig.~\ref{fig:event-sizes}).
Blackouts that are 100 MW and smaller are considered small and the rest are considered large. Blackouts that are 1 day or shorter are considered  considered short and the rest are considered long.
An interesting observation from these figures is that the $n-1$ secure 39 case with no DG has a similar level of total risk to the 39 bus case (original) with 5\% DG on the load buses. However, the long duration and large size events have a larger contribution to risk, relative to the short and large events in the $n-1$ case and are closer to equal in the 39 bus case with 5\% DG. This is presumably related to the fact that the longest recovery time picked from the distribution will always be longer than the recovery time to 99.9\% of the load served for the $n-1$ case.
Note that CRISP assumes the DG is consistently available, which would represent the case of an extremely well-managed DG fleet with a reliable fuel supply.

\section{Conclusion}
This paper presents a new model for measuring power system resilience, known as CRISP (Computing Resilience Interactions Simulation Platform), which is driven by empirical utility data.
The main idea is to represent all the processes of resilience in an integrated way by leveraging both the observed statistics of utility outages and restoration and a (admittedly simple) power grid model. While the individual outage and restoration processes of resilience are not represented in tremendous detail, the observed statistics driving the model do incorporate the effect of the many and detailed mechanisms that contribute to the processes of resilience in the observed statistics. This approach, while approximate, largely avoids the detailed modeling assumptions that limit the strength of conclusions in more detailed simulations. Detailed simulations typically select and sharply approximate a subset of the possible mechanisms of resilience to maintain tractability. We see our comprehensive and high-level resilience simulation as complementary to more detailed simulations, both in assumptions and their application. 

We demonstrate the CRISP model with real data from one utility, but the data is standard and the model is designed so that utilities can easily use their own data to quantify the resilience of their own system.
The paper also illustrates how this model can be used to address important policy questions. In this case we asked and answered the question of whether adding well-managed distributed generation improves the resilience of a power system.
We found that DG significantly improves the resilience of the power system test case we studied. Through testing the original network and the $n-1$ secure network, we found that the first improvement made to the network, whether that be added DG or making the network $n-1$ secure, appears to have the largest effect on improving resilience.

Future improvements to this model will include better models of cascading (spreading) based on empirical data, time-domain changes in load levels, and some modeling of interactions between different types of infrastructure systems.


\begin{thebibliography}{10}

\bibitem{walker2004resilience}
B.~Walker, C.~S. Holling, S.~Carpenter, and A.~Kinzig, ``Resilience,
  adaptability and transformability in social--ecological systems,'' {\em
  Ecology and society}, vol.~9, no.~2, 2004.

\bibitem{liu2005negative}
H.~Liu, R.~A. Davidson, D.~V. Rosowsky, and J.~R. Stedinger, ``Negative
  binomial regression of electric power outages in hurricanes,'' {\em Journal
  of infrastructure systems}, vol.~11, no.~4, pp.~258--267, 2005.

\bibitem{han2009estimating}
S.-R. Han, S.~D. Guikema, S.~M. Quiring, K.-H. Lee, D.~Rosowsky, and R.~A.
  Davidson, ``Estimating the spatial distribution of power outages during
  hurricanes in the gulf coast region,'' {\em Reliability Engineering \& System
  Safety}, vol.~94, no.~2, pp.~199--210, 2009.

\bibitem{Bruneau:Resilience-earthquake}
M.~Bruneau, S.~E. Chang, R.~T. Eguchi, G.~C. Lee, T.~D. O’Rourke, A.~M.
  Reinhorn, M.~Shinozuka, K.~Tierney, W.~A. Wallace, and D.~von Winterfeldt,
  ``A framework to quantitatively assess and enhance the seismic resilience of
  communities,'' {\em Earthquake Spectra}, vol.~19, no.~4, pp.~733--752, 2003.

\bibitem{satumtira2010synthesis}
G.~Satumtira and L.~Due{\~n}as-Osorio, ``Synthesis of modeling and simulation
  methods on critical infrastructure interdependencies research,'' in {\em
  Sustainable and resilient critical infrastructure systems}, pp.~1--51,
  Springer, 2010.

\bibitem{reed17}
S.~Wang and D.~Reed, ``Vulnerability and robustness of civil infrastructure
  systems to hurricanes,'' {\em Frontiers in Built Environment}, vol.~3,
  pp.~2005--2010, 2017.

\bibitem{wu2016modeling}
B.~Wu, A.~Tang, and J.~Wu, ``Modeling cascading failures in interdependent
  infrastructures under terrorist attacks,'' {\em Reliability Engineering \&
  System Safety}, vol.~147, pp.~1--8, 2016.

\bibitem{Dunn18}
S.~Dunn, S.~Wilkinson, D.~Alderson, H.~Fowler, and C.~Galasso, ``Fragility
  curves for assessing the resilience of electricity networks constructed from
  an extensive fault database,'' {\em Natural Hazards Review}, vol.~19, 02
  2018.

\bibitem{MultiElementContScreen:Davis}
C.~M. {Davis} and T.~J. {Overbye}, ``Multiple element contingency screening,''
  {\em IEEE Transactions on Power Systems}, vol.~26, pp.~1294--1301, Aug 2011.

\bibitem{MassiveContAnal:Huang}
Z.~{Huang}, Y.~{Chen}, and J.~{Nieplocha}, ``Massive contingency analysis with
  high performance computing,'' in {\em 2009 IEEE Power Energy Society General
  Meeting}, pp.~1--8, July 2009.

\bibitem{PerfIndex:Ejebe}
G.~C. {Ejebe} and B.~F. {Wollenberg}, ``Automatic contingency selection,'' {\em
  IEEE Transactions on Power Apparatus and Systems}, vol.~PAS-98, pp.~97--109,
  Jan 1979.

\bibitem{MLS:Carleton}
C.~Coffrin, R.~Bent, B.~Tasseff, K.~Sundar, and S.~Backhaus, ``Relaxations of
  ac minimal load-shedding for severe contingency analysis,'' 10 2017.

\bibitem{IanCascade}
H.~Ren, I.~Dobson, and B.~Carreras, ``Long-term effect of the n-1 criterion on
  cascading line outages in an evolving power transmission grid,'' {\em IEEE
  Transactions on Power Systems}, vol.~23, pp.~1217 -- 1225, 09 2008.

\bibitem{Clarfeld2019}
L.~A. {Clarfeld}, P.~{Hines}, E.~{Hernandez}, and M.~{Eppstein}, ``Risk of
  cascading blackouts given correlated component outages,'' {\em IEEE
  Transactions on Network Science and Engineering}, pp.~1--1, 2019.
\newblock early access.

\bibitem{song_dynamic_2016}
J.~Song, E.~Cotilla-Sanchez, G.~Ghanavati, and P.~D.~H. Hines, ``Dynamic
  modeling of cascading failure in power systems,'' {\em IEEE Transactions on
  Power Systems}, vol.~31, pp.~2085--2095, May 2016.

\bibitem{Adibi94}
M.~M. Adibi and L.~H. Fink, ``Power system restoration planning,'' {\em IEEE
  Transactions on Power Systems}, vol.~9, no.~1, pp.~22--28, 1994.

\bibitem{ji2016large}
C.~Ji, Y.~Wei, H.~Mei, J.~Calzada, M.~Carey, S.~Church, T.~Hayes, B.~Nugent,
  G.~Stella, M.~Wallace, {\em et~al.}, ``Large-scale data analysis of power
  grid resilience across multiple us service regions,'' {\em Nature Energy},
  vol.~1, no.~5, p.~16052, 2016.

\bibitem{kwasinski2009telecommunications}
A.~Kwasinski, W.~W. Weaver, P.~L. Chapman, and P.~T. Krein,
  ``Telecommunications power plant damage assessment for hurricane
  katrina--site survey and follow-up results,'' {\em IEEE Systems Journal},
  vol.~3, no.~3, pp.~277--287, 2009.

\bibitem{araneda2010lessons}
J.~C. Araneda, H.~Rudnick, S.~Mocarquer, and P.~Miquel, ``Lessons from the 2010
  chilean earthquake and its impact on electricity supply,'' in {\em 2010
  international conference on power system technology}, pp.~1--7, IEEE, 2010.

\bibitem{coffrin15}
P.~V. Hentenryck and C.~Coffrin, ``Transmission system repair and
  restoration,'' {\em Mathematical Programming}, vol.~151, no.~1, pp.~347--373,
  2015.

\bibitem{Tan2017}
Y.~Tan, F.~Qiu, A.~K. Das, D.~S. Kirschen, P.~Arabshahi, and J.~Wang,
  ``{Scheduling Post-Disaster Repairs in Electricity Distribution Networks},''
  vol.~8950, no.~c, pp.~1--11, 2017.

\bibitem{Panteli:MetricsResilience}
M.~{Panteli}, P.~{Mancarella}, D.~N. {Trakas}, E.~{Kyriakides}, and N.~D.
  {Hatziargyriou}, ``Metrics and quantification of operational and
  infrastructure resilience in power systems,'' {\em IEEE Transactions on Power
  Systems}, vol.~32, pp.~4732--4742, Nov 2017.

\bibitem{Panteli:Fragil-Prob-Adapt}
M.~Panteli, C.~Pickering, S.~Wilkinson, R.~Dawson, and P.~Mancarella, ``Power
  system resilience to extreme weather: Fragility modeling, probabilistic
  assessment, and adaption measures,'' {\em IEEE Transactions on Power
  Systems}, vol.~32, no.~5, pp.~3747--3757, 2017.

\bibitem{Tan2018}
Y.~Tan, A.~K. Das, P.~Arabshahi, and D.~S. Kirschen, ``{Distribution systems
  hardening against natural disasters},'' {\em IEEE Transactions on Power
  Systems}, vol.~33, no.~6, pp.~6849--6860, 2018.

\bibitem{Nan2017}
C.~Nan, G.~Sansavini, W.~Kr{\"{o}}ger, and H.~Heinimann, ``A quantitative
  method for assessing the resilience of infrastructure systems,'' {\em
  Reliability Engineering and System Safety}, vol.~157, pp.~35--53, June 2017.

\bibitem{Alvehag:Reliability-Distrib-Weather}
K.~Alvehag and L.~Soder, ``A reliability model for distribution systems
  incorporating seasonal variations in severe weather,'' {\em IEEE Transactions
  on Power Delivery}, vol.~26, pp.~910 -- 919, 05 2011.

\bibitem{DobsonHICSS18}
I.~Dobson, N.~Carrington, Z.~K., Z.~Wang, C.~B.A., and J.~Reynolds-Barredo,
  ``Exploring cascading outages and weather via processing historic data,'' in
  {\em Fifty-first Hawaii International Conference on System Sciences}, (Big
  Island, Hawaii.), Jan 2018.

\bibitem{DobsonPS12}
I.~Dobson, ``Estimating the propagation and extent of cascading line outages
  from utility data with a branching process,'' {\em IEEE Transactions on Power
  Systems}, vol.~27, no.~4, pp.~2146--2155, 2012.

\bibitem{BPAwebsite}
``{B}onneville {P}ower {A}dministration {T}ransmission {S}ervices {O}perations
  \& {R}eliability website,'' Apr. 2017.
\newblock \url{https://transmission.bpa.gov/Business/Operations/Outages}.

\bibitem{DobsonPS16}
I.~Dobson, B.~Carreras, D.~Newman, and J.~Reynolds-Barredo, ``Obtaining
  statistics of cascading line outages spreading in an electric transmission
  network from standard utility data,'' {\em IEEE Transactions on Power
  Systems}, vol.~31, no.~6, pp.~4831--4841, 2016.

\bibitem{KancherlaPS18}
S.~Kancherla and I.~Dobson, ``Heavy-tailed transmission line restoration times
  observed in utility data,'' {\em IEEE Transactions on Power Systems},
  vol.~33, no.~1, pp.~1145--1147, 2018.

\bibitem{ClausetSIAM09}
A.~Clauset, C.~Shalizi, and M.~Newman, ``Power-law distributions in empirical
  data,'' {\em SIAM Review}, vol.~51, no.~4, pp.~661--703, 2009.

\end{thebibliography}
\end{document}